\documentstyle[prd,floats,aps]{revtex}

\begin{document}

\input{epsf.sty}

\draft

\twocolumn[\hsize\textwidth\columnwidth\hsize\csname
@twocolumnfalse\endcsname

\title{Simple excision of a black hole in 3+1 numerical relativity}

\author{Miguel Alcubierre and Bernd Br\"ugmann}

\address{Max-Planck-Institut f\"ur Gravitationsphysik, 
Albert-Einstein-Institut, Am M\"{u}hlenberg 1, 14476 Golm, Germany}

\date{\today; AEI-2000-034}

\maketitle


\begin{abstract}
  We describe a simple implementation of black hole excision in 3+1
  numerical relativity.  We apply this technique to a Schwarzschild
  black hole with octant symmetry in Eddington-Finkelstein coordinates
  and show how one can obtain accurate, long-term stable numerical
  evolutions.
\end{abstract}

\pacs{04.25.Dm, 04.30.Db, 95.30.Sf, 97.60.Lf}

\vskip2pc]


The simulation of a black hole inspiral collision is one of the
most important open problems facing numerical relativity. Traditional
techniques using singularity avoiding slicings will not be able to
follow such a collision since problems associated with the stretching
of the slice typically cause simulations to crash or to become
extremely inaccurate in time scales far shorter than the orbital time
scale.  Black hole excision techniques (also known as ``apparent horizon
boundary condition''~\cite{Seidel92a,Anninos94e}) appear to be the
most promising way of eliminating the problem of the slice stretching,
thus in principle allowing numerical simulations to follow the
inspiral from well separated holes through the merger and the
ring-down phase.

Black hole excision was first attempted successfully by Seidel and
Suen in spherical symmetry~\cite{Seidel92a}, and was later studied in
more detail by Anninos et.al.~\cite{Anninos94e}. However, the original
idea is older, and Thornburg~\cite{Thornburg87,Thornburg93} has
attributed it to a suggestion by Unruh from 1984.  The idea consists of
two parts: First, one places a boundary inside the black hole and
excises its interior from the computational domain.  Second, one uses
a shift vector that keeps the horizon roughly in the same coordinate
location during the evolution (``horizon tracking'',
see~\cite{Thornburg93}).  Since no information can leave the interior
of the black hole, excision should have no effect on the physics
outside. Ideally, one would like to know the position of the event
horizon which marks the true causal boundary, but the global character
of its definition means that in principle one can only locate it once the whole
evolution of the spacetime is known.  The apparent horizon, on the
other hand, can be located on every time slice and is guaranteed to be
inside the event horizon.  In practice one therefore needs to find the
apparent horizon and excise a region contained inside it.

Though black hole excision has been successful in spherical
symmetry~\cite{Seidel92a,Anninos94e,Scheel94,Marsa96,Gundlach98a,Scheel97,Scheel98a,Kidder00a},
it has been difficult to implement with a 3+1 approach in
three-dimensions
(3D)~\cite{Anninos94c,Daues96a,Bruegmann96,Walker98a}, where
instabilities typically plague the evolutions (but some progress has
been made, see~\cite{Cook97a,Thornburg99}).  Black hole excision using
a characteristic formulation, on the other hand, has been very
successful in 3D, allowing stable evolutions of perturbed black holes
for thousands of $M$'s~\cite{Gomez98a}.  However, such characteristic
formulations are likely to have problems with the development of
caustics in the case of extremely distorted or colliding black holes,
so the search for a stable 3+1 excision implementation is still of
great importance.

Here we present a 3+1 approach to black hole excision in 3D that has
allowed us to obtain long-term stable, accurate evolutions of a single
black hole spacetime.  These results are currently limited to
simulations in octant symmetry as discussed below.


\medskip

\begin{center}
{\bf I. SIMPLE BLACK HOLE EXCISION}
\end{center}

Though conceptually simple, black hole excision in 3D is a complicated
problem numerically.  First, one has to cut a hole in the
computational domain that has a spherical topology and is therefore
not well adapted to the Cartesian coordinates typically used.  Second,
one has to apply some condition at the boundary of the excised region
that is stable and respects the causality of the physical system.  As
the excised region is inside a black hole, no boundary condition
should be needed since all the information required to update the
boundary comes from outside the excised region.  However, achieving
this ``boundary without a boundary condition''
(BWBC)~\cite{Seidel92a,Gundlach98a} in 3D is difficult, particularly
if one uses a formulation of the evolution equations that is not
hyperbolic.  The way this problem is usually approached is by using
``causal differencing''~\cite{Seidel92a,Anninos94e} or ``causal
reconnection''~\cite{Alcubierre94a}, where the computational molecules
are adapted to follow the causal structure.  The mixture of these
issues makes it difficult in practice to identify what particular
element of an algorithm is responsible for causing a numerical
simulation to go unstable.

In our approach we have simplified the algorithm as much as possible,
separating out what we believe is essential to the excision problem.
Our algorithm is based on the following simplifications:

\begin{itemize}
  
\item Excise a region adapted to Cartesian coordinates, i.e.
  excise a cube contained inside the horizon.
  
\item Do not attempt to fulfill the BWBC ideal, and use instead a
  simple but stable boundary condition at the excision boundary.
  
\item Do not use causal differencing.  Use instead centered
  differences in all terms except the advection terms on the shift
  (terms that look like $\beta^i \partial_i \,\,$).  For these terms
  use upwind along the shift direction (we use the standard 1D
  second-order upwind stencil in each of the Cartesian coordinate
  directions based on the sign of the corresponding shift component at
  each point).  This is very important, as it is the only place where
  any information about causality (i.e. the direction of the shift)
  enters our scheme.  Using a centered approximation for these terms
  results in an unstable scheme.

\end{itemize}

One can worry that excising a cube will introduce artifacts into the
evolution, but as long as the boundary condition used at the sides of
the cube is consistent those artifacts will converge away with
increased resolution.  Similarly, one can argue that applying a
boundary condition instead of using causal differencing is
inconsistent with the physics, but since this condition is applied
well inside the horizon, any error introduced is unlikely to propagate
outside the hole.


\medskip

\begin{center}
{\bf II. STATIC BLACK HOLE SPACETIME}
\end{center}

As the first test of our excision algorithm we have considered a
single static black hole written in ``3+1 Eddington-Finkelstein''(EF)
coordinates. These 3+1 EF coordinates are a simple transformation of
the standard ingoing EF coordinates~\cite{Misner73} to a 3+1 form.
The resulting metric has no coordinate singularities, penetrates the
event horizon, reaches the physical singularity, and is manifestly
time independent.  This makes it ideal for excision tests where one
can excise the physical singularity and try to keep the numerical
evolution stable and close to static. The 3+1 EF metric has the form
\begin{eqnarray}
ds^2 &=& - \left( 1 - 2 M/r \right) dt^2 + \left( 4 M /r \right)
dt dr \nonumber \\
&& + \left( 1 + 2 M/r \right) dr^2 + r^2 d\Omega^2 \; .
\label{eq:ourEF}
\end{eqnarray}
with $M$ the black hole mass and $d\Omega$ the solid angle element.
From this metric one can read the values of the 3-metric, lapse and
shift.  The extrinsic curvature can then be obtained in a
straightforward way.


\medskip

\begin{center}
{\bf III. EVOLUTION EQUATIONS}
\end{center}


{\em Formulation}--- We comment briefly on the formulation used for
the simulations described below.  Our simulations have been performed
using a formulation of the 3+1 evolution equations developed by
Baumgarte and Shapiro~\cite{Baumgarte99} (BS), based on previous work
of Shibata and Nakamura~\cite{Shibata95} (SN). The motivation for
using this BSSN formulation comes from the fact that it has shown
remarkable stability properties when compared to the
Arnowitt-Deser-Misner (ADM) formulation~\cite{Arnowitt62} in a wide
range of numerical
simulations~\cite{Baumgarte99,Baumgarte99b,Shibata99c,Shibata99d,Shibata99e,Alcubierre99b,Alcubierre99e}.

The BSSN variables are defined in terms of the spatial metric
$\gamma_{ij}$ and the extrinsic curvature $K_{ij}$ as: \mbox{$\phi =
  \mbox{ln}(\mbox{det} \gamma_{ij})/12$}, \mbox{$\tilde\gamma_{ij} =
  e^{-4\phi} \gamma_{ij}$}, \mbox{$\mbox{tr}K=\gamma^{ij} K_{ij}$},
\mbox{$\tilde A_{ij} = e^{-4\phi}(K_{ij}-\gamma_{ij} {\rm tr}K/3)$},
and \mbox{$\tilde\Gamma^i = \tilde\gamma^{jk} \Gamma^i_{jk}$} (note
that \mbox{$\mbox{det} \tilde g = 1$} and $\mbox{tr}\tilde A = 0$).
See~\cite{Baumgarte99} for the explicit form of the evolution
equations, and~\cite{Alcubierre99e} for an analysis that indicates why
the BSSN formulation should be superior to ADM at least for linearized
perturbations of flat space.

In order to obtain the stable evolutions described below, we have
found it necessary to add the following ingredients to the BSSN
formulation:

\begin{enumerate}
  
\item As discussed in~\cite{Alcubierre99d}, we actively force the
  trace of the conformal-traceless extrinsic curvature
  $\tilde{A}_{ij}$ to remain zero during our simulations by
  subtracting it after each time step.
  
\item We use the independently evolved ``conformal connection
  functions'' $\tilde\Gamma^i$ only in terms where derivatives of
  these functions appear.  Whenever these functions are
  undifferentiated, we recompute them from the conformal Christoffel
  symbols.  We have found this to be very important to achieve long-term
  stability, but at the moment we lack a theoretical understanding as
  to why this is so.

\end{enumerate}


{\em Slicing conditions}--- As a first approach to evolving the
solution described above, one could think of using the exact value of
the lapse.  It turns out that it is difficult to keep the evolution
stable if the lapse is not allowed to adapt to the (numerically
induced) evolution of other dynamical quantities, particularly the
trace of the extrinsic curvature.  In order to obtain stable
evolutions we have found it crucial to use a ``live'' slicing
condition.  What is required is a slicing condition that is well
adapted to the exact solution in the sense that for this solution it
recovers the exact lapse.  For this we start from the Bona-Mass\'{o}
family of slicing conditions~\cite{Bona94b}
\begin{equation}
\partial_t \alpha = - \alpha^2 \, f \left(\alpha \right) {\rm tr} K \, ,
\label{eq:BMlapse}
\end{equation}
with $f(\alpha) > 0$.  As it is, this condition does not reproduce our
exact solution for which ${\rm tr}K \neq 0$, but $\partial_t
\alpha$=0.  However, one can easily see that for zero shift
Eq.~(\ref{eq:BMlapse}) implies $\partial_t \alpha \propto \partial_t
(\mbox{det}g)$.  For this to hold also with non-zero shift
Eq.~(\ref{eq:BMlapse}) must be generalized to
\begin{equation}
\partial_t \alpha = - \alpha \, f \left(\alpha \right) \left[ \alpha
\, {\rm tr} K - \nabla_i \beta^i \right] \, .
\label{eq:ourlapse}
\end{equation}
For any static solution Eq.~(\ref{eq:ourlapse}) implies
$\partial_t \alpha$=0.

Another natural slicing condition to consider is $\partial_t {\rm tr} K$=0. 
For initial data with ${\rm tr}K$=0 this condition leads to
maximal slicing, but $\partial_t {\rm tr} K$=0 is a gauge choice
that can be made in general, even if ${\rm tr}K \neq 0$, as is the case
for the constant time slices of the black hole in EF coordinates. This
``K freezing'' condition leads to an elliptic equation for
the lapse,
\begin{equation}
  \Delta \alpha - \alpha K_{ij} K^{ij} - \beta^i \nabla_i {\rm tr} K 
  = 0.
\label{eq:Kfreezing}
\end{equation}
In the numerical implementation, we solve this equation for the lapse
but we hold $\mbox{tr} K$ constant in time by hand.  In
\cite{Alcubierre99b} in the context of the evolution of strong waves
we have found that otherwise a drift away from the initial value due
to numerical errors can lead to an instability.  Such drifts were one
of the reasons that led us to consider trace-split formulations like
BSSN, because here $\mbox{tr} K$ is evolved as an independent
variable which makes it trivial to enforce $\partial_t\mbox{tr} K = 0$.


{\em Shift conditions}--- In contrast to the experience with the
lapse, we have found that using a static (exact) shift does allow us
to get long-term stable evolutions.  However, this is not useful in
general, so we have considered also live shift conditions. Live
shifts have been studied before for black hole spacetimes
in~\cite{Daues96a}, where a minimal distortion shift
condition~\cite{Smarr78b} led to limited stability ($t \sim 100M$) for
a single excised black hole.

In our case a good choice was a conformal version of the 3-harmonic
shift~\cite{Smarr78a}. 3-harmonic shifts play a natural role in mixed
elliptic-hyperbolic systems~\cite{Andersson98}.  The condition we
impose in the BSSN system is $\partial_t \tilde\Gamma^k$=0 (``Gamma
freezing'' condition, note that $\tilde\Gamma^k \neq 0$), or
\begin{eqnarray}
  \tilde\gamma^{jk} \partial_j\partial_k \beta^i
+ \frac{1}{3} \tilde\gamma^{ij}  \partial_j\partial_k\beta^k
- \tilde\Gamma^j \partial_j \beta^i 
+ \frac{2}{3} \tilde\Gamma^i \partial_j\beta^j
+ \beta^j\partial_j \tilde\Gamma^i
&& \nonumber \\
- 2 \tilde A^{ij} \partial_j\alpha
- 2 \alpha \left( \frac{2}{3}\tilde\gamma^{ij}\partial_j {\rm tr} K 
- 6 \tilde A^{ij}\partial_j\phi - \tilde \Gamma^i_{jk} \tilde A^{jk} \right)
 = 0 .
&& \nonumber \\
&& \hspace*{-3cm}
\label{eq:Gammafreezing}
\end{eqnarray} 
As mentioned before, $\partial_k \tilde\Gamma^i$ is computed from the
independent variable $\tilde\Gamma^i$, in other terms we use
$\tilde\Gamma^i = \tilde\gamma^{jk} \tilde\Gamma^i_{jk}$ (notice that
the momentum constraint was used to replace $\partial_j\tilde A^{ij}$
in~(\ref{eq:Gammafreezing})). Equation~(\ref{eq:Gammafreezing}) is an
elliptic equation for the shift vector. For the solution of
(\ref{eq:Kfreezing}) and (\ref{eq:Gammafreezing}) we have used the
multi-grid solver from BAM~\cite{Bruegmann97}.
As in the case of K freezing, we explicitly hold the value of
$\tilde\Gamma^i$ constant in time in order to prevent this quantity
from drifting due to numerical errors. As shown in Section IV, allowing
$\tilde\Gamma^i$ to drift results in an unstable evolution.

We have also looked at shift prescriptions given by evolution
equations instead of elliptic conditions.  One way to do this is to
transform an elliptic equation into a parabolic one by making
$\partial_t \beta^i$ proportional to the given elliptic operator
(``driver'' conditions, see~\cite{Balakrishna96a}).  As an example we
considered the following evolution equation for the shift obtained
from the Gamma freezing condition (a ``Gamma driver'' condition)
\begin{equation}
\partial_t \beta^i = k \, \partial_t \tilde\Gamma^i \, ,
\qquad (k > 0) \, .
\label{eq:Gammadriver}
\end{equation}


{\em Boundary conditions}--- There are two very different boundaries
to consider in our simulations: the outer boundary of the numerical
grid, and the inner boundary of the excised region.  In principle
there should be a rigorous treatment of numerical boundaries at finite
radii (starting e.g.\ from~\cite{Friedrich99}, the first analytic
treatment of the initial boundary value problem). Here we are looking
for simple numerical methods that are sufficient for the evolution of
excised black holes.

At the outer boundary we have attempted to keep all fields equal to
their exact values, but have found that this introduces late time
instabilities. Using a live boundary condition allows us to eliminate
these instabilities.  The boundary condition we use is a radiative
boundary condition applied to the difference between a given variable
and its exact value: \mbox{$f-f_{\rm exact}=u(r-t)/r$}.  We apply this
condition to all fields (even to the lapse and shift in the case of
the algebraic gauge conditions) {\em except}\/ the $\tilde{\Gamma}^i$
which we leave fixed to their exact values at the boundary.  Applying
this condition to the $\tilde{\Gamma}^i$ causes a drift away from the
exact solution that eventually crashes the simulation (the origin of
this drift is not well understood, but it seems to be related to the
shift choice and is not present if one uses the Gamma driver shift
described above).

As to what boundary condition to use at the sides of the excision
cube, we have experimented with many different conditions and have
finally settled on one that simply copies the time derivative of every
field at the boundary from its value one grid-point out along the
normal direction to the cube (at edges and corners we define the
normal direction as the diagonal).  This condition is perfectly
consistent with evolving a static solution, where the time derivatives
are supposed to be zero.  Even in a dynamical situation, this condition
is still consistent with the evolution equations since it is
equivalent to just calculating the source term one grid point away.
This means that our boundary condition should introduce a first order
error, but as mentioned above, we do not expect this error to affect the
solution outside the horizon.  One could in principle argue that
nothing prevents gauge modes and constraint violating modes from
propagating outside the horizon, thus spoiling the second order
convergence of the exterior scheme.  We have looked carefully at the
convergence of our simulations, and have found no evidence that this
happens in practice.


\medskip

\begin{center}
{\bf IV. NUMERICAL RESULTS}
\end{center}

We now present some results of our numerical simulations.  As
discussed above, our simulations have been done with a live lapse
condition, and we have considered both a static shift, and several
live shift conditions.  In our runs we have always taken $M$=1, so the
horizon is a sphere of radius $r$=2, and we excise a cube of side~1
(we have in fact also excised cubes of different size, but the results
discussed below are not affected by this).  The numerical integration
is carried out using the so-called iterative Crank-Nicholson scheme
with 3 iterations (counting the initial Euler step as iteration 1).
Because of the spherical symmetry of the problem typically only one
octant was evolved (with positive $x$, $y$, and $z$). However, as
discussed at the end of this section, an unstable mode appears when
the same simulations are performed on the corresponding full grids.


\medskip

\begin{figure}
\epsfxsize=3.5in
\epsfysize=2.2in
\epsfbox{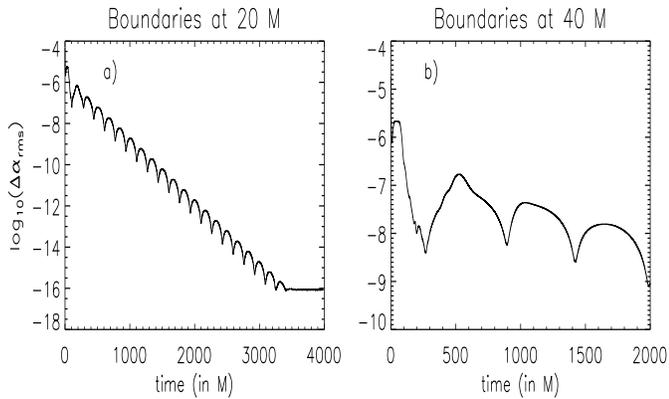}
\caption{Log plot of r.m.s.\ of the change in the lapse;
  $\Delta x = 0.4$, $\Delta t = 0.1$. a) $53^3$ grid points, boundary
  at $20M$. b) $103^3$ grid points, boundary at $40M$.}
\label{fig:exact_dtalp}
\end{figure}

{\em Static shift}--- We first consider the case when the shift
remains equal to its exact value.  Figure~\ref{fig:exact_dtalp} shows
a log plot of the root mean square (r.m.s.) of the change in the lapse
between consecutive time steps for two simulations using slicing
condition~(\ref{eq:ourlapse}) with $f = 1/\alpha$ (``1+log''
slicing~\cite{Alcubierre99d,Arbona99}), a grid spacing $\Delta x =
0.4$, and a time step $\Delta t = 0.1$. Figure~\ref{fig:exact_dtalp}a
shows the results of a simulation using $53^3$ grid points, with the
outer boundaries at $20M$.  The change in the lapse drops as an
exponentially damped oscillation until at $t\sim 3500 M$ it reaches
the level of round-off error ($10^{-16}$) and settles down (other
functions show a similar behavior). The evolution was stopped at
$t=4000 M$, but it is clear that it could have continued.
Figure~\ref{fig:exact_dtalp}b shows a simulation with the same
resolution, but using $103^3$ grid points, with the outer boundaries
now at $40M$.  The simulation goes past $t \sim 2000 M$, and seems to
have settled on an exponentially decaying oscillating pattern. (This
simulation took 100 hours running on 16 processors of an Origin 2000
SGI machine.  If the pattern continues, round-off error level would be
reached by $t \sim 12000 M$, requiring another $500 \times 16$ hours
of computer time).  The most obvious differences between the run with
the boundaries at $20M$ and that with the boundaries at $40M$ is the
fact that the period of the oscillations increases and the rate of
decay decreases.  The period increases by a factor of 3.4 as we double
the distance to the outer boundaries, so the oscillation time scale is
not given directly by the light travel time from the boundary (which
would approximately double).  We do not know exactly what fixes this
time scale, but the fact that when we look at individual metric components
we see that the oscillations behave like standing waves (and not
travelling pulses) would seem to indicate that we are looking at
different modes of oscillation of the whole system (interior plus
boundaries).

\begin{figure}[t]
\epsfxsize=3.5in
\epsfysize=2.2in
\epsfbox{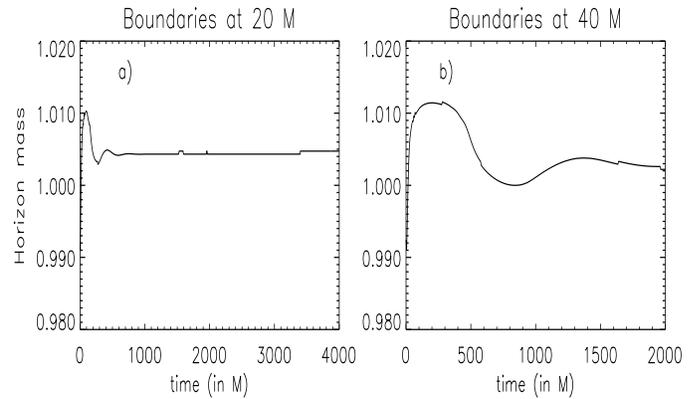}
\caption{Evolution of horizon mass for the same simulations.}
\label{fig:exact_mass}
\end{figure}

These simulations are not only stable for very long times, they are
also exceedingly accurate.  We have located the apparent horizon every
50 time-steps (using the 3D finder described in~\cite{Alcubierre98b}),
measured its area $A$ and computed its mass $M = \sqrt{A/(16 \pi)}$.
Figure~\ref{fig:exact_mass} shows the behavior of the horizon mass as
a function of time.  In both cases, after an initial transient, the
mass settles on a stationary value with an error of less than $1\%$.

\begin{figure}[t]
\epsfxsize=3.5in
\epsfysize=2.2in
\epsfbox{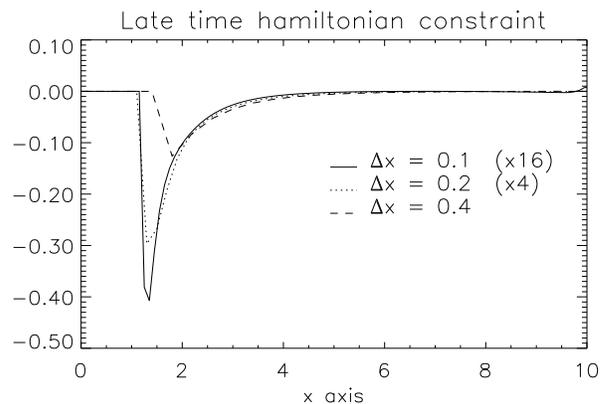}
\caption{Late time Hamiltonian constraint for runs with different
  resolutions.  The values for the higher resolution runs were
  multiplied by factors of 4 and 16.}
\label{fig:exact_conv}
\end{figure}

In Figure~\ref{fig:exact_conv} we consider the convergence of our
simulations by looking at the late time value of the Hamiltonian
constraint along the $x$ axis for simulations with $28^3$, $53^3$ and
$103^3$ grid points and resolutions of $\Delta x$=0.4,0.2,0.1
respectively (boundaries at $10M$).  The Hamiltonian constraint
for the higher resolution runs has been multiplied by factors of 4 and
16.  The fact that the three lines coincide indicates second order
convergence.


\medskip

\begin{figure}
\epsfxsize=3.4in
\epsfysize=2.5in
\epsfbox{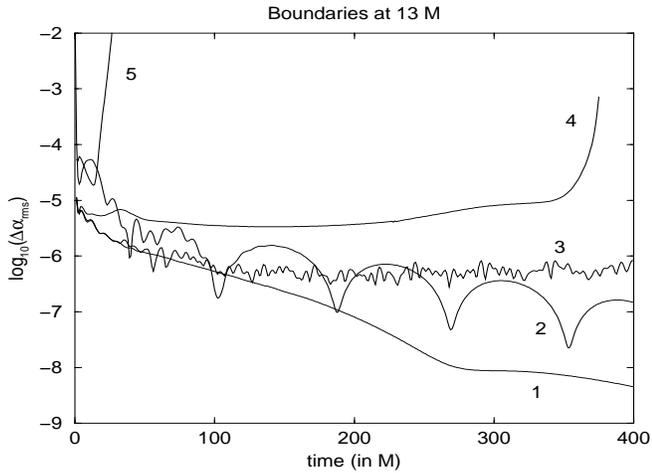}
\caption{Log plot of r.m.s.\ of the change in the lapse 
  for different lapse and shift combinations involving elliptic
  conditions; $\Delta x = 0.4$, $\Delta t = 0.1$, $35^3$ points,
  boundary at 13$M$.  Run~1: stable ($\Gamma$ freezing without drift,
  1+log); run~2: stable ($\Gamma$ freezing without drift, K freezing
  without drift); run~3: unstable ($\Gamma$ freezing without drift,
  1+log, static outer boundaries); run~4: unstable ($\Gamma$ freezing
  with drift, 1+log); run~5: unstable (minimal distortion, 1+log).}
\label{fig:srlf_ham}
\end{figure}

\begin{figure}[t]
\epsfxsize=3.4in
\epsfysize=2.5in
\epsfbox{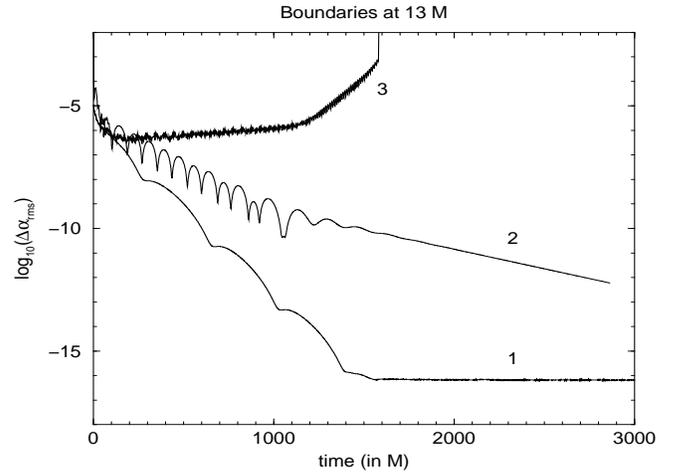}
\caption{Runs 1, 2, and 3 of Figure~\ref{fig:srlf_ham} for run times
of up to $t = 3000M$.}
\label{fig:srlf_ham2}
\end{figure}

{\em Elliptic shifts}--- We now consider results with elliptic shifts,
such as those that we expect will be needed in a 3D black hole
merger simulation.  Figure~\ref{fig:srlf_ham} shows two stable and
three unstable runs up to $t = 400M$, and Figure~\ref{fig:srlf_ham2} shows
those three runs that lasted longer 
up to $t = 3000M$. Second order convergence has been checked using two
grids with $19^3$ and $35^3$ points with the outer boundary at
$7M$. For 1+log slicing a radiative boundary
condition is applied to the lapse, while lapse and shift for the
elliptic conditions are held fixed at the exact values.

Stable runs are obtained for Gamma freezing shift with either 1+log or
K freezing slicings. Referring to Figures \ref{fig:srlf_ham} and
\ref{fig:srlf_ham2}, for 1+log slicing $\Delta\alpha_{{\rm rms}}$
falls below $10^{-16}$ at $t \sim 1500M$ after four oscillations (run 1),
while for K freezing there are more than fifteen oscillations, which
damp out at around $10^{-10}$ followed by a straight line decay (run 2).

The 1+log, Gamma freezing run becomes unstable if the boundary values
of all fields are static (run 3, crashing at $t \sim 1500M$,
Figure~\ref{fig:srlf_ham}b), or if $\partial_t\tilde\Gamma^i = 0$ is
not set to zero identically and is allowed to drift because of
numerical errors (run 4, crashing at $375M$). We also tested 1+log
slicing with a minimal distortion shift~\cite{Smarr78b} computed from
the ADM variables, but this run fails already at $27M$ (run 5).


\medskip

\begin{figure}
\epsfxsize=3.5in
\epsfysize=2.2in
\epsfbox{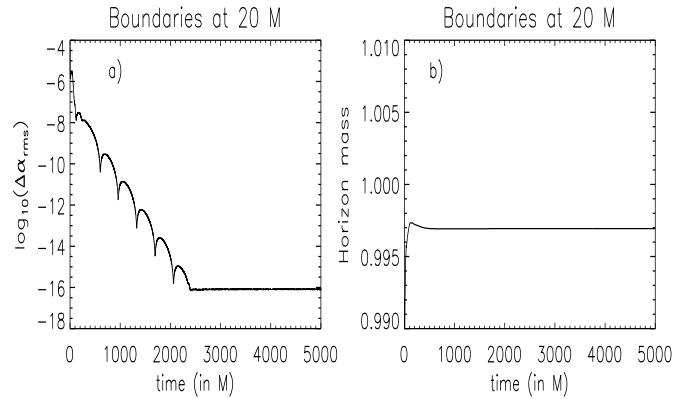}
\caption{Simulation using Gamma driver shift with $k$=0.1; $\Delta x = 0.4$,
  $\Delta t = 0.1$, $53^3$ points, boundary at 20$M$. a) Log plot of
  r.m.s. of change in the lapse. b) Horizon mass.}
\label{fig:driver}
\end{figure}

{\em Algebraic shifts}--- Finally, we consider a simulation using
1+log slicing and a Gamma driver shift with $k=0.1$.
Figure~\ref{fig:driver} shows the r.m.s.\ of the change in the lapse
and the horizon mass for a simulation with $\Delta x = 0.4$, $\Delta t
= 0.1$ and $53^3$ grid points.  After $t \sim 2500M$ the solution
becomes static up to round-off error.

\medskip

\begin{figure}
\epsfxsize=3.5in
\epsfysize=2.2in
\epsfbox{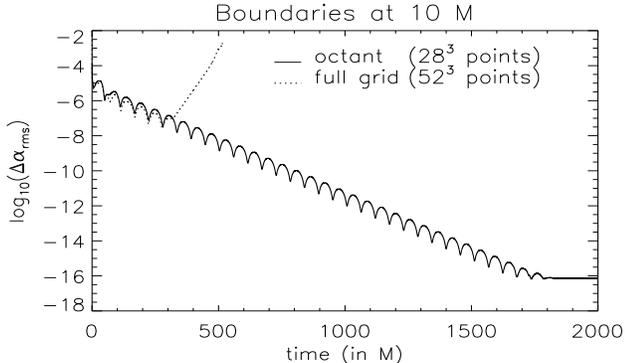}
\caption{
  Unstable mode on a full grid for 1+log slicing with a static shift.
  Shown is a log plot of the r.m.s.\ of the change in the lapse for an
  octant run with $\Delta x = 0.4$, $\Delta t = 0.1$, and $28^3$ grid
  points together with the corresponding full grid run.}
\label{fig:full}
\end{figure}

{\em Discussion}--- The above results demonstrate that stable 3D black
hole runs can be obtained with the simple excision technique that we
introduced in this paper, with a variety of different gauge
conditions. However, repeating these runs on a full grid as opposed to
just one octant, with otherwise identical parameters, uncovers an
unstable mode. Figure~\ref{fig:full} shows as an example the situation
for 1+log slicing and static shift, although the problem appears for
all the gauge conditions considered here.  Tracing the growth of the
unstable mode back in time suggests that it has started as numerical
round-off error of around $10^{-14}$ at $t=0$.  Increasing the grid
resolution appears to have no significant effect on the growth rate of
the unstable mode, but the simulation now crashes slightly sooner.
However, we do see good second order convergence at early times, before
the instability becomes apparent. The situation does not improve if
we impose the exact data at the excision boundary (imposing exact data
at the excision boundary in octant mode works well and leads to stable
simulations).  Also, the presence of a horizon does not seem to be the
cause of the problem since when we excise a cube that contains the
horizon, as opposed to being contained by it, the instability is still
present although it becomes somewhat milder (not surprising since we
have excised a region with stronger data).  While the achievable run
times of about $500M$ are roughly 10 times larger than for singularity
avoiding slicings, we have found that introducing an artificial
asymmetry on the full grid by simply off-setting the excision box one
grid point in all directions makes the runs fail much sooner.
Although the slope of the blow-up is not significantly affected when
this artificial asymmetry is introduced, the exponential growth
becomes evident from the very beginning.  On the other hand, the full
grid runs can be stabilized by setting certain terms in the BSSN
equations to their analytic values.  In particular, freezing the
evolution of the $\tilde \Gamma^i$ while keeping the shift static
suffices to obtain stability.  In conclusion, the instability appears
to be more directly linked to the system of evolution equations than
to the boundary condition, and we will investigate different
variations of the evolution system in the future.

We have also repeated the above simulations using the ADM equations
with the same gauge and boundary conditions, and the same numerical
techniques, but these runs fail already at $t \simeq 30M$ even in
octant mode.


\medskip

\begin{center}
{\bf V. CONCLUSIONS}
\end{center}

We have described a black hole excision technique in 3+1 numerical
relativity that has allowed us to obtain accurate, long-term stable
evolutions of a black hole spacetime in 3D. The main limitation is that
the transition from octant symmetry to full grids introduces an
unstable mode, which is currently under investigation. Our
implementation of excision is based on the idea of simplifying all
ingredients of the excision algorithm as much as possible. In our case
this means: 1) excising a cube naturally adapted to the underlying
Cartesian coordinates, 2) imposing a simple but stable boundary
condition on the sides of this cube, and 3) using an upwind scheme
instead of causal differencing.  Crucial for obtaining our long-term
stable evolutions has been the use of a live slicing condition and a
radiation outer boundary condition.  Although keeping a static shift
does not appear to have a detrimental effect on the stability of our
simulations, we have also experimented with several live shift
conditions, both algebraic and elliptic, that can be generalized to
more interesting physical situations. We consider these results a
necessary first step towards the development of excision techniques
capable of evolving the full inspiral collision of two black holes in
an accurate and stable way.


\acknowledgements We would like to thank J.~Baker, D.~Pollney,
E.~Seidel, W.-M.~Suen and J.~Thornburg for useful discussions and
comments. The numerical experiments were implemented using BAM and the
Cactus Computational Toolkit~\cite{Bruegmann99b,Cactusweb}. All
computations were performed at the Max-Planck-Institut f\"ur
Gravitationsphysik.


\bibliographystyle{prsty}
\bibliography{bibtex/references}

\end{document}